\documentclass{JHEP3}
\pdfoutput=1 
\usepackage[english]{babel}
\usepackage{amsmath,amssymb}
\usepackage[pdftex]{graphicx}
\usepackage[sort&compress]{natbib}
\usepackage{amsfonts}
\usepackage{bbm}

\newcommand\fverb{\setbox\fverbbox=\hbox\bgroup\verb}
\newcommand\fverbdo{\egroup\medskip\noindent%
			\fbox{\unhbox\fverbbox}\ }
\newcommand\fverbit{\egroup\item[\fbox{\unhbox\fverbbox}]}
\newbox\fverbbox

\newcommand{\eg} {{\it e.g.}}
\newcommand{\ie} {{\it i.e.}}

\newcommand{\bc}{\begin{center}}
\newcommand{\ec}{\end{center}}
\newcommand{\bt}{\begin{tabular}}
\newcommand{\et}{\end{tabular}}
\newcommand{\be}{\begin{equation}}
\newcommand{\ee}{\end{equation}}
\newcommand{\bea}{\begin{eqnarray}}
\newcommand{\eea}{\end{eqnarray}}
\newcommand{\bfig}{\begin{figure}}
\newcommand{\efig}{\end{figure}}

\def\gsim{\mathrel{\lower3pt\hbox{$\sim$}}\hskip-11.5pt\raise3pt\hbox{$>$}\;}

\def\lsim{\mathrel{\lower3pt\hbox{$\sim$}}\hskip-11.5pt\raise3pt\hbox{$<$}\;}

\title{IDM \& iDM\\or \\ The Inert Doublet Model and Inelastic Dark Matter}

\author{Chiara Arina, Fu-Sin Ling, Michel H.G. Tytgat\\Service de Physique Th\'eorique, Universit\'e Libre de Bruxelles,
  CP225, Bld du Triomphe, 1050 Brussels, Belgium\\E-mail: \email{carina@ulb.ac.be}, \email{fling@ulb.ac.be}, \email{mtytgat@ulb.ac.be}}

\preprint{ULB-TH/09-21}

\abstract{The annual modulation observed by DAMA/NaI and DAMA/Libra may be interpreted in terms of elastic or inelastic scattering of dark matter particles. 
In this paper we confront these two scenarios within the framework of a very simple extension of the Standard Model, the Inert Doublet Model (IDM). 
In this model the dark matter candidate is a scalar, the lightest component of an extra Higgs doublet. 
We first revisit the case for the elastic scattering of a light scalar WIMP, $M_{DM} \sim 10$ GeV, 
a scenario which requires that a fraction of events in DAMA are channelled. 
Second we consider the possibility of inelastic Dark Matter (iDM). 
This option is technically natural in the IDM, in the sense that the mass splitting between the lightest and next-to-lightest neutral scalars 
may be protected by a Peccei-Quinn (PQ) symmetry. 
We show that candidates with  a mass $M_{DM}$ between $\sim 535$ GeV and $\sim 50$ TeV may reproduce the DAMA data and have a cosmic 
abundance in agreement with WMAP. 
This range may be extended to candidates as light as $\sim 50$ GeV if we exploit the possibility  that
the approximate PQ symmetry is effectively conserved and that a primordial asymmetry in the dark sector may survive until freeze-out.}


\begin{document}

\section{Introduction}\label{sec:intro}

DAMA/Libra, and the former DAMA/NaI are dark matter (DM) direct detection experiments (DAMA in the sequel) which have observed 11 successive cycles 
of annual modulation in the rate of nuclear recoils, with a statistical significance of $8.2 \sigma$ \cite{Bernabei:2008yi}. 
These measurements are consistent with the signal that would arise from collisions between nuclei and weakly interacting massive particles (WIMP) 
from a galactic halo, the flux of dark matter particles being  modulated by the 
periodic motion of the Earth around the Sun~\cite{Drukier:1986tm,Freese:1987wu}. 
For a standard, Maxwellian, spherical, non-rotating halo, the collision rate is expected to peak around the 2nd of June, 
when the velocity of the Earth with respect to the halo of dark matter adds to the velocity of the Sun in the Galaxy.

A problem with DAMA is that a straightforward interpretation of the data in terms of WIMP collisions  
is in contradiction with the exclusion limits set by other direct detection experiments 
(see for instance \cite{Gelmini:2008vi} or \cite{Hooper:2009zm} for  recent reviews). 
Ways  to reconcile the DAMA observations with the other experiments include models of mirror dark matter \cite{Foot:2003iv,Foot:2008nw}, 
composite dark matter \cite{Khlopov:2008ki}, MeV dark matter \cite{Bernabei:2008mv}, hidden charged dark matter \cite{Feng:2009mn}, 
dipolar dark matter \cite{Masso:2009mu}, etc. In the present work, we focus on two alternative scenarios. 
Concretely, we consider the spin-independent (SI) elastic scattering of a light WIMP, with $M_{DM}$ in the 
multi-GeV range \cite{Gelmini:2004gm,Gondolo:2005hh,Bottino:2002ry,Bottino:2003iu,Bottino:2003cz,
Bottino:2007qg,Petriello:2008jj,Chang:2008xa,Fairbairn:2008gz,Savage:2008er,Savage:2009mk,Dudas:2008eq,Andreas:2008xy,Feng:2008dz}, 
and the so-called Inelastic Dark Matter (iDM) 
scenario \cite{TuckerSmith:2001hy,TuckerSmith:2004jv,Chang:2008gd,MarchRussell:2008dy,Cui:2009xq,Finkbeiner:2009ug}. 
To a large extent, our conclusions concur with those obtained in these works, but our main intention is distinct, 
as we wish to confront a specific particle physics model, the so-called Inert Doublet Model (or IDM), to the recent DAMA data. 

The IDM is a very simple extension of the Standard Model, first introduced in~\cite{Deshpande:1977rw} 
but which has received some attention more recently, starting in \cite{Barbieri:2006dq,Ma:2006km}. 
Despite its simplicity, the IDM has an interesting phenomenology, 
which may be related to the origin of neutrino masses \cite{Ma:2006km} or to electroweak symmetry 
breaking \cite{Barbieri:2006dq,Hambye:2007vf} and various aspects of dark matter have been 
discussed~\cite{LopezHonorez:2006gr,Gustafsson:2007pc,Andreas:2008xy,Agrawal:2008xz,Andreas:2009hj,Nezri:2009jd,Dolle:2009fn}. 
Also, the IDM is one instance of a whole category of models for dark matter, to which models like 
Minimal Dark Matter \cite{Cirelli:2005uq} or Scalar Multiplet Models \cite{Hambye:2009pw} belong.

The IDM contains two Higgs doublets and,
to prevent FCNC, a discrete $Z_2$ symmetry is imposed, with the extra (or inert) Higgs  taken to be odd under $Z_2$. 
All the other fields, with the possible exception of right-handed neutrinos~\cite{Ma:2006km}, are even. 
There are two extra neutral scalars states, noted $H_0$ and $A_0$,  which, if $Z_2$ is unbroken, may play the role of dark matter. 
Here the lightest state, hence the dark matter candidate, is taken to be $H_0$. 
The mass splitting  between these two states is controlled by a quartic coupling to the Higgs, here $\lambda_5$, with 
$$
M_{H_0}^2 - M_{A_0}^2 = \lambda_5 v^2
$$
where $v = 246$ GeV is the vacuum expectation value of the Higgs. 
In the limit $\lambda_5 \rightarrow 0$, the $Z_2$ symmetry is elevated to a global $U(1)_{PQ}$ Peccei-Quinn symmetry, 
a limit  relevant for the inelastic dark matter scenario, which requires a small splitting, $M_{H_0} - M_{A_0} \sim 100$ keV. 
In this limit, scattering and annihilation are dominantly through the $Z$ channel. 
Another interesting limit is that of a light $H_0$ and two heavier, nearly degenerate states, {\em i.e.} $A_0$ and the charged 
partner in the inert doublet. This limit, which is relevant for elastic scattering of a light WIMP scenario, 
is also protected by a global  symmetry, which is $SU(2)$ in this case, analog of the custodial symmetry of the Standard Model. 
In this scenario, scattering and annihilation are dominantly through the Higgs channel. 
The elastic and inelastic scenarios for DAMA correspond to specific limits in the parameter space of the IDM.
The main purpose of this paper is to discuss the behavior of the model in these two specific regions, in connection 
with the present direct detection experiments.

The organization of the paper is as follows. The IDM is introduced in Section \ref{sec:IDM}. 
The physics of direct detection, including the seasonal modulation, and the relevant data are discussed in Section \ref{sec:dde}.  
We discuss the elastic scenario in Section \ref{sec:elwimp}. The main result of this section is summarized in Figure \ref{fig:mlbda},  
in which we compare the goodness of fit of a light WIMP candidate in IDM to the DAMA data, together with 
the exclusion limits set by the other experiments and with the WMAP relic abundance. 
At $3 \sigma$, only a small region of the parameter space is compatible with all observations. 
We then discuss the inelastic scenario in Section \ref{sec:inelwimp}. 
The main result of this section is summarized in Figure~\ref{fig:mdelta}, 
in which we show that a whole range of candidates between $M_{DM} \sim 535$ GeV and $M_{DM}\sim 50$ TeV is 
compatible with all observations, including the WMAP abundance. 
In the same section, we argue that lower mass candidates, which in the standard freeze-out scenario have a too small relic abundance, 
may be consistent with all observations, provided there is a primordial  PQ asymmetry in the dark sector. 
We give our conclusions in Section \ref{sec:conclusions}. 

\section{The Inert Doublet Model}\label{sec:IDM}

The Inert Doublet Model (IDM) is a simple extension of the Standard Model with dark matter, with two Higgs doublets and a $Z_2$ symmetry.  
The usual Brout-Englert-Higgs doublet (hereafter the Higgs doublet) is denoted by $H_1$.
The extra, or inert doublet, $H_2$, is the only field of the model that is odd under the $Z_2$ symmetry.\footnote{Actually, this is not the only logical possibility. Since fermions come in pairs, as one may also impose that all the fermions of the Standard Model are odd under $Z_2$, without jeopardising the stability of the neutral scalar \cite{Kadastik:2009dj,Kadastik:2009cu}. As emphasized in these works, this so-called 'matter-parity' naturally points to a scalar dark matter candidate and furthermore opens the possibility to embed the model in SO(10).} 
This ensures the stability of the lightest member of $H_2$, which will be a DM candidate, and prevents from flavor
changing neutral currents (FCNC)~\cite{Deshpande:1977rw}. 
We will assume that $Z_2$ is not spontaneously broken and that $H_2$ does not develop a vacuum expectation value. 
We write $H_2=(H^+ \quad (H_0+iA_0)/\sqrt{2})^T$,
similarly to the ordinary Higgs doublet, and $H_1=(h^+ \quad (v_0+h+iG_0)/\sqrt{2})^T$. The potential 
is written as
\begin{equation}\label{potential}
\begin{split}
V(H_1,H_2) &= \mu_1^2 \vert H_1\vert^2 + \mu_2^2 \vert H_2\vert^2  + \lambda_1 \vert H_1\vert^4
+ \lambda_2 \vert H_2\vert^4 \\
&+ \lambda_3 \vert H_1\vert^2 \vert H_2 \vert^2 + \lambda_4 \vert H_1^\dagger H_2\vert^2
+ {\lambda_5\over 2} \left[(H_1^\dagger H_2)^2 + h.c.\right]\,.
\end{split}
\end{equation}
 After electroweak symmetry breaking, 
$\langle H_1 \rangle = v = -\mu_1^2/\lambda_1 = 246$~GeV, the masses of the physical scalar fields are given by
\bea
M_h^2 &=& 2 \lambda_1 v^2~,\cr
M_{H_0}^2 &=& \mu_2^2 +  \lambda_{H_0} v^2~,\cr
M_{A_0}^2 &=& \mu_2^2 +  \lambda_{A_0}  v^2~,\cr
M_{H^+}^2 &=& \mu_2^2 + \lambda_{H_c} v^2~,
\label{masses}
\eea
with $\lambda_{H_c} \equiv \lambda_3/2$ and
$\lambda_{H_0,A_0} \equiv (\lambda_3 + \lambda_4 \pm
\lambda_5)/2$. We will consider $H_0$ to be the DM candidate
(\ie~$\lambda_5<0$) though the results would be exactly the same
for $A_0$ changing the sign of $\lambda_5$.
Notice that most of the negative couplings parameter space is excluded by vacuum stability constraints. 
Indeed, to ensure that the scalar potential is bounded from below, we need \cite{Barbieri:2006dq}
\bea
\lambda_{1,2} &>& 0 \quad , \nonumber \\
\lambda_{H0}~, \quad \lambda_{A_0} \, , \quad \lambda_{H_c} &>& - \sqrt{\lambda_1\lambda_2}~.
\label{vacstab}
\eea

The IDM has already been extensively studied in the literature. It has been shown
that a viable DM candidate with the correct relic abundance can be obtained in
three regimes, low-mass ($M_{H_0} \ll m_W$) \cite{Hambye:2007vf,Andreas:2008xy},
middle-mass ($M_{H_0} \lsim m_W$)\cite{Barbieri:2006dq,LopezHonorez:2006gr} and
high-mass ($M_{H_0} \gg m_W$)\cite{LopezHonorez:2006gr,Hambye:2009pw}.
Direct and indirect detection constraints were investigated in
Refs.~\cite{Barbieri:2006dq,Majumdar:2006nt,LopezHonorez:2006gr,Gustafsson:2007pc,
Andreas:2008xy,Agrawal:2008xz,Andreas:2009hj,Nezri:2009jd} and confrontation to colliders data and related 
future prospects was done in~\cite{Cao:2007rm,Lundstrom:2008ai}. Here we will focus on the implications of DAMA for the IDM.

Two particular limits of the mass relations will be most relevant here. 
The most obvious one is the limit in which $\lambda_5 \rightarrow 0$, in which case the neutral particles are degenerate, 
$M_{H_0} = M_{A_0}$, and the theory is invariant under a larger, 
Peccei-Quinn symmetry,  $U(1)_{PQ} \supset Z_2$. This limit will be relevant in Section \ref{sec:inelwimp} 
where we will discuss the inelastic scenario for DAMA. The existence of an enhanced symmetry will imply that the 
limit of nearly degenerate neutral scalars, albeit fine-tuned,  is technically natural. 
From Eqs.~(\ref{masses}), a small mass splitting $\delta = M_{A_0}-M_{H_0} \simeq 100$~keV as suggested by DAMA corresponds to a coupling
\be
\lambda_5 = 3.3 \cdot 10^{-7} \left( \frac{M_{H_0}}{100~\rm{GeV}} \right) \left( \frac{\delta}{100~\rm{keV}} \right) \, .
\label{lambdacinq}
\ee
It is worth noticing that the IDM is the only model among Scalar Multiplet Dark Matter models, Ref.~\cite{Hambye:2009pw},
where such a tiny mass splitting 
between the neutral components can be generated at the renormalizable level. This opens the interesting possibility that 
for higher scalar multiplets the splitting may be generated by higher order operators. 

Another, perhaps less obvious limit is that with $M_{A_0} = M_{H^\pm}$, in which case three real states are 
degenerate and the theory is invariant under a global $SU(2)$ symmetry \cite{Gerard:2007kn,Hambye:2007vf}. 
This symmetry is analogous to the custodial symmetry in the Standard Model, in the sense that it prevents the 
appearance of large radiative corrections to the $\rho$, or equivalently $T$, parameter.   
This implies that, to some extent, it is possible to decouple the extra components of the inert doublet, $M_{H_0} \ll M_{A_0} \approx M_{H^\pm}$.  
In this limit, the dark matter phenomenology is  similar to that of a singlet scalar field \cite{Andreas:2008xy}, 
a model for dark matter much discussed in the literature \cite{McDonald:1993ex,Burgess:2000yq,Barger:2007im}.

\section{Direct detection experiments}
\label{sec:dde}
For the sake of reference, we give all the relevant steps necessary to compute the signal due to SI scattering of 
dark matter on nuclei in direct detection experiments. A most useful reference is the review by Lewin and 
Smith \cite{Lewin:1995rx}. See also Jungman, Griest and Kamionkowski~\cite{Jungman:1995df}.  

\subsection{Event rate}
We consider collisions between dark matter particles from the Galactic halo and the nuclei of a given low background detector. 
The relevant characteristics of the halo are the local density of dark matter, 
taken to have the fiducial value $\rho_{DM}=0.3$ GeV/cm$^{3}$ at the Sun's location, and the local  distribution of velocities ${\rm f}(\vec v)$
with respect to the Earth.   

At a given recoil energy $E_R$, the differential event rate of nuclear recoils can be factored as
\begin{equation}
\label{eq:diffrate}
\frac{{d}R}{{d}E_{R}} = \frac{\rho_{DM}}{M_{DM}} \frac{{d}\sigma}{{d}E_{R}} \; \eta (E_R,t)
\end{equation}
where $M_{DM}$ is the WIMP mass, 
${d}\sigma/{d}E_{R}$ encodes all the particle and nuclear physics factors, and
$\eta (E_R,t)$ is the mean inverse velocity of the incoming particles that can deposit a given recoil energy $E_R$. 
The time dependence of the velocity distribution is induced by the motion of the Earth around the Sun, which leads to a 
seasonal modulation of the event rate \cite{Drukier:1986tm,Freese:1987wu}. 

The total recoil rate per unit detector mass in a given energy bin $[E_1,E_2]$ is obtained by integrating Eq.~(\ref{eq:diffrate}),
\begin{equation}
\label{eq:totrate}
R(t) = \int_{E_1}^{E_2} {d}E\  \epsilon(E)\  \left( \frac{{d}R}{{d}E} \star G(E,\sigma(E)) \right) \quad ,
\end{equation}
where $\epsilon$ is the efficiency of the detector and the finite energy resolution of the experiment is taken into account by
convoluting the differential rate with a gaussian distribution with spread $\sigma(E)$. 
For detectors made of several elements, the total rate is the average of the rates $R_i(t)$ for each component $i$, 
weighted by its mass fraction $f_i$
\begin{equation}
R(t) = \sum_{i} f_i R_i(t)
\end{equation}
Finally the expected number of observed events per unit time is the product of the total rate times the detector mass $M_{det}$. 
We will discuss further effects, like 
quenching factors and channelling effects in the next subsections, together with the other characteristics of a specific experiment.

The particle physics is enclosed in the term ${d}\sigma/{d}E_{R}$, which is generally parametrized as
\begin{equation}
\label{cross_section}
\frac{{d}\sigma}{{d}E_{R}} = \frac{M_N \sigma^0_n}{2 \mu^2_n}\ \frac{\Big(f_p^2 Z + (A-Z) f_n^2\Big)^2}{f_n^2} F^2(E_R) \quad ,
\end{equation}
where $M_N$ is the nucleus mass, $\mu_n$ is the reduced WIMP/neutron mass, $\sigma^0_n$ is the zero momentum WIMP-neutron effective cross-section, 
$Z$ and $A$ are respectively the number of protons and the atomic number of the element, 
and $f_p$ ($f_n$) are the WIMP effective coherent coupling to the proton (resp. neutron). 
The nuclei form factor $F^2(E_R)$ characterizes the loss of coherence for non zero momentum transfer.
We use the simple parametrization given by Helm~\cite{Helm:1956zz,Lewin:1995rx}, defined as
\begin{equation}
\label{eq:F_Iodine}
F(E_R) = 3 e^{- q^2 s^2/2}\ \frac{\sin(qr)- qr\cos(qr)}{(qr)^3}
\end{equation} 
with $q=\sqrt{2 M_N E_R}$, $s=0.9$ fm and $r$ the effective nuclear radius. 
This form factor is optimal for scattering on Iodine \cite{Lewin:1995rx}. 
For simplicity we use the same form factor for all targets (more accurate form factor are off by at most ${\cal O}(20 \%)$, 
for some targets and at large recoil energies \cite{Duda:2006uk}).  

Finally, the velocity distribution appears in the quantity
\begin{equation}
\eta (E_R,t) =  \int_{v_{min}} {\rm d}^3 \vec{v}\  \frac{ {\rm f}(\vec{{v}}(t))}{{v}}  \quad ,
\end{equation}
where $\vec{v}$ the WIMP velocity {\it wrt} the Earth and $v_{min}$ is the minimum velocity needed to provoke a recoil inside the detector. 
The threshold velocity is given by
\be
v_{min} = \sqrt{\frac{1}{2 M_N E_R}} \Big(\frac{M_N E_R}{\mu}+\delta\Big) \quad ,
\ee
with $\mu$ the nucleus-WIMP reduced mass. This formula  encompasses both  elastic ($\delta$ = 0) and inelastic ($\delta \neq 0$) scatterings.
Here $\delta$ is the mass splitting between the two DM particles involved in the scattering 
(which are different in the case of inelastic scatterings).

When the velocity distribution in the galactic frame ${\rm f_{gal}}(v)$ is isotropic, $\eta$ is given by
\be
\eta = \frac{2\pi}{v_\oplus} \int_{v_-}^{v_+} \big( {\rm F}(v_{esc}) - {\rm F}(v) \big) \, dv \quad ,
\label{myeta}
\ee
with $v_\pm = \min\{v_{esc},v_{min} \pm { v}_{\oplus}\}$, $v_{esc}$ is the galactic escape velocity,
$\vec{v}_\oplus(t) = \vec{v}_\odot + \vec{v}_{{\rm EO}}(t)$ is the Earth's velocity in the galactic frame, 
with $\vec v_\odot = (0,220,0) + (10,13,7)$ km/s is the velocity of 
the Sun {\em wrt} the halo~\cite{Gelmini:2000dm}, $v_{\rm EO} = 29.8$~km/s is the Earth mean orbital velocity, 
and ${\rm F}(v)=\int v \, {\rm f_{gal}}(v) \, dv$. Notice that ${\rm F} (v)$ is an even function of $v$ because ${\rm f}_{{\rm gal}}(v)$ only depends on the modulus of the velocity. In deriving this formula, we have used the realistic assumption 
$v_{esc} > v_\oplus$ at any time $t$.
For a standard Maxwellian distribution,
\be
{\rm f_{gal}}(\vec{v}) = \frac{1}{N(v_{esc})\pi^{3/2}v_0^3} \, e^{-v^2/v_0^2} \quad {\rm for}~v<v_{esc} \quad , 
\ee
where $N(v_{esc})$ is a normalization factor, Eq.~(\ref{myeta}) reads
\be
\eta = \frac{1}{2N(v_{esc})v_\oplus} \left\{ {\rm Erf}\Big( \frac{v_+}{v_0} \Big) - {\rm Erf}\Big( \frac{v_-}{v_0} \Big) 
- \frac{2(v_+-v_-)}{\sqrt{\pi} \, v_0} \, e^{-v_{esc}^2/v_0^2}  \right\} \quad ,
\ee
which agrees with the result of Ref.~\cite{Savage:2008er}. It might seem curious that Eq.~(\ref{myeta}) is expressed
as a definite integral between $v_-$ and $v_+$ while particles with any velocity above $v_{min}$ should contribute to $\eta$.
However, the function $F(v)$ is itself an integral on the velocity distribution. Also the dependence in $v_{esc}$ is explicit.

In this paper, all the fits and $\chi^2$ analyses have been performed with a standard Maxwellian halo velocity distribution with
a dispersion parameter $v_0=220~\rm{km/s}$, and an escape velocity $v_{esc}=450$ and $v_{esc}=600$ km/s or $v_{esc} =650$ km/s, 
somewhat on the edges of the expected range, $498\  \mbox{\rm km/s}\,< v_{esc}< 608 \,\mbox{\rm km/s}$ \cite{Smith:2006ym}. 
Modifications of the velocity distributions and their impact on the fits, have been discussed more in depth in various works, 
see for instance Refs.~\cite{Fairbairn:2008gz,Savage:2009mk,MarchRussell:2008dy,Vergados:2007nc}.

\subsection{DAMA}\label{subsec:dama}
The former DAMA/NaI \cite{Bernabei:2000qi} and present DAMA/LIBRA \cite{Bernabei:2008yi} experiments are made of NaI(Tl) crystals. 
They are designed to detect the dark matter recoil off nuclei through the model independent annual modulation signature 
due to the motion of the Earth around the Sun.  The experimental results obtained by
DAMA/LIBRA, with an exposure of 0.53 ton$\times$yr collected over 4 annual cycles, combined with the ones of DAMA/NaI, 
for an exposure of 0.29 ton$\times$yr collected over 7 annual cycles, corresponding to a total exposure of 0.82 ton$\times$yr, 
show a modulated signal with a confidence level of 8.2 $\sigma$~\cite{Bernabei:2008yi}. 

For an isotropic velocity distribution, the contribution of the signal to the counting rate in each bin can be approximated as
\begin{equation}
\frac{{d}R}{{d}E_{R}} \simeq S_0(E_R) +S_m(E_R) \cos \omega (t-t_0)
\end{equation}
where $S_0$ is the average signal, $S_m$ is the modulation amplitude, $\omega=2\pi/T$ with T = 1 year, and $t_0$ 
is the phase corresponding to $t_0=152.5$ day (June 2), the time of year at which $v_{\oplus}(t)$ is at its maximum. 
The amplitude of the modulated signal is therefore given by:
\begin{equation}\label{eq:sm}
{ S_
  m (E_R}) = \frac{{d}R}{{d}E_{R}}\Big|_{mod} \simeq \frac{1}{2}\ \Big\{\frac{{d}R}{{d}E_{R}}(\mbox{\rm June} \ 2)-
  \frac{{d}R}{{d}E_{R}}(\mbox{\rm December} \ 2)  \Big\} \,.
\end{equation}
DAMA gives the modulated average amplitude over bin intervals, with
\begin{equation}\label{eq:smbin}
S_m = \frac{1}{E_2-E_1} \int_{E_1}^{E_2} {d}E\  S_m(E)\,.
\end{equation}
The energy resolution of the detector given by the collaboration~\cite{Bernabei:2008yi} is
\begin{equation}
\frac{\sigma(E)}{E}= \frac{0.45}{\sqrt{E}}+0.0091 
\end{equation}
and the efficiency is $\epsilon =1$ \cite{Bernabei:2008yi,Savage:2008er}. 

The DAMA spectrum is given in  keVee (electron equivalent eV). 
The observed energy released in scintillation light is  related to the nuclei recoil energy through a so-called 
quenching factor $q$, $E_{\mbox{\rm scint}} = q \cdot E_{\mbox{\rm recoil}}$. This expresses the fact that a recoiling 
nucleus may loose energy by collisions with other nuclei, hence in the form of heat, or through collisions with electrons, 
which create scintillation light.  
The reference values for Iodine and Sodium are respectively $q_I = 0.09$ and $q_{Na} = 0.3$. 
However it has been pointed that so-called channelled events may play a role \cite{Bernabei:2007hw,Bottino:2007qg}. 
This refers to events in which a recoiled nucleus moves along the axis of the NaI crystal, 
losing most of its energy by collisions with electrons, in which case the quenching factor may be larger, up to $q\approx 1$.  
Once channelling is taken into account, collisions of light WIMPs with Iodine become relevant, 
while recoils on sodium are negligible in all 
scenarios \cite{Bottino:2007qg,Bottino:2008mf,Petriello:2008jj,Savage:2008er,Fairbairn:2008gz,Chang:2008xa}.  
We  use the fraction $f$ of channelled events given in Ref.~\cite{Fairbairn:2008gz}, 
\begin{equation}
f_{\rm Na}(E_R) = {e^{-E_R/18}\over 1+ 0.75 E_R}\,, \qquad f_{\rm I} = {e^{-E_R/40}\over 1 + 0.65 E_R}\,
\end{equation}
where the recoil energy $E_R$ is in keV. We have verified that the other parametrizations that we have found in the literature, 
\cite{Chang:2008gd,Petriello:2008jj,Savage:2008er}, give identical results.

We compare the prediction of the IDM  to the observed energy spectrum of Eq.~(\ref{eq:sm}), 
using all the data provided by DAMA, which are given in Figure 9 of Ref.~\cite{Bernabei:2008yi}, 
in  36 bins of width $0.5$ keVee, from $2.0$ to $20.0$ keVee~\cite{Bernabei:2008yi}. 
For the explicit numerical values, we refer to the Table III of Ref.~\cite{Savage:2008er}. 

We use a basic statistical analysis. 
First we keep all the 36 bins, even though most of the signal is supposed to be concentrated around the first twelve to eighteen bins. 
This leads to some dilution of the statistical significance of the signal, which in turns means that we are more 
optimistic about which parameters of our model may actually fit the data. However our results are essentially consistent 
with those of \cite{Savage:2008er,Fairbairn:2008gz,Chang:2008xa,Chang:2008gd,MarchRussell:2008dy}\footnote{There is no consensus regarding how to bin the DAMA data. The analysis of \cite{Petriello:2008jj} for instance is based on the published number of events given by the collaboration \cite{Bernabei:2008yi}, which are combined in two bins (2-6 keVee and 6-14 keVee). The analysis of \cite{Savage:2008er} is based on the 36 bins given in the figure 9 of Ref. \cite{Bernabei:2008yi}, recombined in the first 16 bins plus 1 bin from 10-20 keVee. The analysis of \cite{Fairbairn:2008gz} uses all the 36 bins, while \cite{MarchRussell:2008dy,Cui:2009xq} keep only the first twelve bins. Here we use all the available data, like in Ref. \cite{Fairbairn:2008gz}, with the reserve expressed in text above.}.   

 We use a Goodness-of-Fit (GOF) method with a $\chi^2$ metric to determine the allowed regions at 90\%, 99\% and $99.9\%$ of confidence level (CL), with
\begin{equation}
\chi^2 = \sum_{i=1}^{n}\frac{(S_i-S_i^{obs})^2}{\sigma^2_i}
\end{equation}
where $S_i$ are the theoretical predictions and $S_i^{obs}$ the reported signals in each bin and $\sigma_i$ 
are the experimental uncertainties in the measurements. For 36 bins and a scan over two parameters the 90\% (99\% and $99.9\%$) CL 
corresponds to $\chi^2 < 45$ (resp. $\chi^2<56$ and $\chi^2< 65$).

We have made numerous comparisons with the other existing analyses, and have found
 good agreement \cite{Savage:2008er,Fairbairn:2008gz,Chang:2008xa,Chang:2008gd,MarchRussell:2008dy}. 
 As a further check, we have also repeated the above procedure considering only a two bins version of the 
 data (\ie\,  the actual numbers published by DAMA) for 2-6 keVee and 6-14 keVee bins, as in 
 Refs.~\cite{Fairbairn:2008gz,Petriello:2008jj,Savage:2008er}. As is well-known, the allowed parameter regions are 
 much larger but in the present work, so as to avoid the cluttering of the figures,  we only show our results using  the complete spectral data. 

As emphasized in Refs.~\cite{Fairbairn:2008gz,Petriello:2008jj,Savage:2008er}, the total unmodulated rate provides a 
further constraint on the DM parameters, we have compared the total 
experimental rate from~\cite{Bernabei:2008yi} with the unmodulated spectrum in each bin, and required that the predicted 
total number of events do not exceed the measured signal.

\subsection{Exclusion limits}\label{subsec:exclusion}

So far all the other direct detection experiments searching for dark matter are compatible with null results. 
In this section we briefly describe the experiments that lead to the most constraining limits on both the elastic, 
Section~\ref{sec:elwimp}, and the inelastic scenarios, Section~\ref{sec:inelwimp}. 

Light nuclei, like Aluminium and Silicon, are more sensitive to light WIMPs scattering $M_{DM} \sim $ multi-GeV, 
and provide the strongest upper bounds on the allowed parameter space favoured by DAMA. In the inelastic scenario, 
which involves heavier candidates, experiments made of heavy nuclei, like Iodine, Xenon and Tungsten are the most constraining ones. 
Germanium made detectors  fall in-between. 

In computing the rate of Eq.~(\ref{eq:totrate}) for each experiments, 
we  uniformly assume that the small number of events seen, if any, are signals from dark matter and we use the Poisson 
statistics to find the parameter space excluded at a given confidence level \footnote{More sophisticated statistical methods exist to set exclusion limits, which are especially powerful in the presence of few events and of an unknown background, like the maximum gap, described in Ref. \cite{Yellin:2002xd}. However they are not systematically adopted in the recent theoretical analyses to which we refer in this paper. In the literature, Poisson statistics has been used in the analysis of Refs. \cite{Fairbairn:2008gz,Cui:2009xq}, while Ref. \cite{MarchRussell:2008dy} and Ref. \cite{Petriello:2008jj}  use the maximum gap method. The most sophisticated analysis is that of Ref. \cite{Savage:2008er}, where they use a mixture of techniques. The results of these analyses are however consistent with each others. For the sake of simplicity, and because we are more focused on the features of our model than on the statistical analysis, we use standard Poisson statistics. We have checked that our results are consistent with those of the other analyses. We have also verified that using the maximum gap method affects only slightly our exclusion limits and that our conclusions regarding the viability of the model are essentially unchanged.}. Integrating over the exposure time, 
the upper bounds are obtained by requiring that the total number of events $N_{tot}$  is compatible with the number of observed 
events at $99\%$ of CL, as in Ref.~\cite{Cui:2009xq}, which means that we are optimistic about the occurrence of signals for dark matter. 
We have checked that our simple procedure reproduces fairly well the published 90\% CL upper bounds in the plane $M_{DM}-\sigma^0_n$ plane, \eg \ Refs. \cite{Fairbairn:2008gz,Savage:2008er,Chang:2008gd}. 
In the following we describe the main features of the considered experiments, and specify when we use a different 
approach to obtain the exclusion limits.\\

\underline{CDMS}: \  The Cryogenic CDMS experiment at Soudan Underground Laboratory operates Ge and Si made solid-state detectors. 
For a heavy WIMP, the Ge data are more constraining: we have considered the ensemble of the three released runs, 
with respectively an exposure of 19.4 kg-day~\cite{Akerib:2004fq}, 34 kg-day~\cite{Akerib:2005kh} after cuts and a total 
exposure of 397.8 kg-days before cuts from~\cite{Ahmed:2008eu}. The searches on Si, released in~\cite{Akerib:2005kh}, 
are more sensitive to light DM candidates, with  a total exposure of 65.8 kg-day before cuts. 
The sensitivity to nuclear recoils is in the energy range between 10-100 keV while the efficiencies and the energy resolution 
of the detectors are given in~\cite{Savage:2008er}. Therefore in the case of the Germanium detectors the efficiency is parametrized as $\epsilon(E_R)=0.25+0.05(E_R-10\  \rm{keV})/5\ \rm{keV}$ for $10\ \rm{keV}<E_R<15\ \rm{keV}$ and $\epsilon(E_R)=0.30$ for the remaining sensitivity range. The total efficiency for the Silicium detectors is given by $\epsilon(E_R) = 0.80 \times 0.95 (0.10+0.30 (E_R-5\ \rm{keV})/15\ \rm{keV})$ for $5\ \rm{keV}<E_R< 20\ \rm{keV}$ and $\epsilon(E_R)= 0.80 \times 0.95 (0.40+0.10 (E_R-20\ \rm{keV})/80\ \rm{keV})$ for $20\ \rm{keV}<E_R<100\ \rm{keV}$. Our conservative exclusion limits are obtained by requiring a total number 
of events $N_{tot}$ less than 8.4, compatible with zero observed events at $99\%$ confidence level.
 
\underline{XENON10}: \   Xenon10 is a dual-phase Xenon chamber operating at LNGS. 
The collaboration has published the data analyses in Refs.~\cite{Angle:2007uj,Angle:2008we}. 
The total exposure is 316.4 kg-days and 10 candidate events have been seen in the recoil energy range between 4.5-26.9 keV. 
In this case we proceed with a different analysis, as in Ref.~\cite{Fairbairn:2008gz}, using the 7 bins as well the bin dependent efficiencies provided by the 
collaboration in Table I of~\cite{Angle:2007uj}, and a $\chi^2$ for Poisson distributed data. 
The constraints are given by imposing the GOF be compatible with zero observed events at 99\% CL. 
This experiment is sensitive to inelastic dark matter in a similar way as the DAMA experiment, 
due to the close proximity of the target nucleus masses.

\underline{ZEPLIN-III}: \   The ZEPLIN-III experiment~\cite{Lebedenko:2008gb} in the Palmer Underground Laboratory at Boulby 
is a two-phase Xenon chamber, so is more sensitive to heavy DM particles. 
The 2008 run has a total exposure of 126.7 kg-day and has seen 7 events in the measured energy range ($E_q$) between 2-16 keVee. 
The quenching factor for the Xenon is parameterized as in~\cite{MarchRussell:2008dy}. 
It is energy dependent for  recoil energies up to 10 keVee, $q_{Xe}=(0.142 E_q + 0.005) \exp{[-0.305 E_q^{0.564}]}$ 
and then becomes constant with a value of $q_{Xe} = 0.48$. The recoil energy $E_R$ is obtained by rescaling $E_R=E_q/q_{Xe}$. 
The exclusion at 99\% CL is obtained by imposing $N_{tot}< 16$.\footnote{The former ZEPLIN-II experiment is similar to ZEPLIN-III. 
ZEPLIN-II had a total exposure of 225 kg-day and 29 events were observed.  
In the analysis of~\cite{Chang:2008gd,Cui:2009xq,MarchRussell:2008dy} these events are accounted as being possible signals of dark matter. 
The recent work of Ref.~\cite{Cline:2009xd} however claims that all the events are consistent with being background, 
which obviously makes the constraint from ZEPLIN-II more stringent. 
In analogy with the Feldamn and Cousins statistics~\cite{Feldman:1997qc} they used, 
here we have considered a Poisson statistics compatible with at most 10 events
which gives exclusion limits similar from those derived in \cite{Cline:2009xd}. 
However we do not reach the same conclusions regarding the constraints on inelastic dark matter, 
see Section~\ref{sec:inelwimp}. }  

\underline{CRESST}: \ For the CRESST experiments at LNGS, we have derived the exclusion limits in both the light and the heavy WIMPs scenarios. 
For a light dark matter we have used the first data release (hereafter CRESST-I) for the prototype $\rm{Al}_2 \rm{O}_3$ detector 
module runs~\cite{Angloher:2002in}, with a total exposure of 1.51 kg-days covering the energy range between 0.6 - 20 keV, 
with 11 observed events and an energy resolution given by $\sigma(E) = \sqrt{ (0.519\  keV)^2+(0.0408\  E)^2}$. 
The eleven  events after cuts produce an upper bound on the total number of events of 23 at 99\% CL. 
The collaboration has changed significantly the detectors for the second commissioned run of 
2007~\cite{Angloher:2004tr,Angloher:2008jj} (CRESST-II) using $\rm{CaWO}_4$ as single crystals. 
The presence of the heavy nucleus of Tungsten enhances the sensitivity to spin-independent inelastic particle scattering. 
The energy range of the CRESST-II Zora and Verena detectors is 12-100 keV, with a total exposure of 47.9 kg - day 
before cuts and an acceptance of 0.9 on Tungsten recoil. The collaboration measured 7 events, 
that provides a limit on the total rate in the modules at 99\% CL compatible with $N_{tot}<16$.\\

Two other experiments are potentially relevant, CoGeNT and TEXONO. 
The CoGeNT experiment~\cite{Aalseth:2008rx} is based on Ge sub-keV threshold crystal detectors. 
The binned data arise from the first prototype of 0.475 kg Ge with a 8.4 kg-days exposure, 
in the measured energy range between 0.388-0.983 keVee, with a quenching factor $q_{Ge}=0.2$. 
The TEXONO experiment~\cite{Lin:2007ka,Avignone:2008xc} runs at the Kuo-Sheng(KS) 
Laboratory and is again a Ge crystal with low threshold, 0.2 keVee. The experiment is sensitive in the energy range 
between 0.2 -0.8 keVee, with the same quenching factor as CoGeNT; the detector has a total exposure of 0.338 kg-days 
with an efficiency of 50\%. 
The limits set by these two experiments are only relevant for a two bins analysis of the DAMA 
data~\cite{Petriello:2008jj,Chang:2008xa,Fairbairn:2008gz,Savage:2008er}, but
to avoid the cluttering of the figures, we do not show them in this paper.

\section{Elastic scattering of a light WIMP}\label{sec:elwimp}
\begin{figure}
\begin{minipage}[t]{0.6\textwidth}
\centering
\includegraphics[width=0.9\columnwidth]{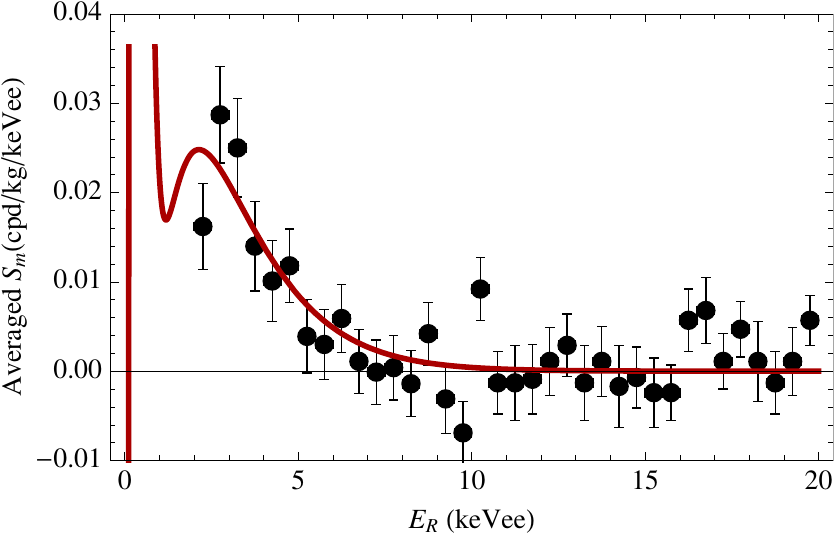}
\end{minipage}
\hspace*{-0.2cm}
\begin{minipage}[t]{0.4\textwidth}
\vspace*{-5cm}
\centering
\begin{tabular}{|cc|}
\hline
& \\
\includegraphics[width=0.25\columnwidth]{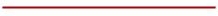}  & $M_{H_0}= 12$ GeV \\
&  $|\lambda_{H_0}| = 0.24$ \\
&   $\chi^2 = 36.6$ \\ 
&  $v_{esc}=650$ km/s \\
&   \\
\hline
\end{tabular}
\end{minipage}
\caption{\underline{Elastic scenario}\ \ The averaged modulated amplitude $S_{m}$ for the best fit point (see Figure~\ref{fig:mlbda}). 
The relevant parameters are given in the table on the right. The $\chi^2_{min}$ is evaluated for 34 d.o.f.}
\label{fig:chi_el}
\end{figure}
The light WIMP scenario refers to a dark matter particle with a mass in the multi-GeV range. 
This is lighter than a typical WIMP candidate, like a standard neutralino, which would have a mass in the multi-10 GeV range. 
The possibility of explaining the DAMA results with a light WIMP has been first considered in  \cite{Gelmini:2004gm,Gondolo:2005hh} and, 
for neutralino candidates, in \cite{Bottino:2002ry,Bottino:2003iu,Bottino:2003cz}. These early proposals, 
which made use of the lower threshold for scattering of light dark matter particles on Sodium rather than, 
say, on Germanium, have been excluded by more recent experimental results. However, the possible effect of 
channelling in the NaI crystals of DAMA \cite{Drobyshevski:2007zj,Bernabei:2007hw}, which effectively lowers 
the threshold for detection of nuclear recoils and reinforces the impact of collisions on Iodine, 
has reopened the possibility of explaining  the data with a light WIMP \cite{Bottino:2007qg}. 
The impact of channelling has been  further studied in many recent 
works \cite{Petriello:2008jj,Bottino:2008mf,Chang:2008xa,Fairbairn:2008gz,Savage:2008er,Savage:2009mk}. 
A general conclusion of these works is that a light WIMP candidate may give a honest fit to the DAMA data, 
but is only marginally consistent with current exclusion limits, with only a small region in the mass vs (SI) 
cross section parameter space remaining  at the 3$\sigma$ level. A further issue of a light WIMP scenario is 
that the total rate of nuclear recoils tends to be large and raising at low recoil energies, 
features that are absent in the data and which furthermore imply that background in DAMA is unexpectedly 
small \cite{Fairbairn:2008gz,Savage:2008er}. Despite these caveats, there is much appeal to a light WIMP candidate. 
For instance,  it is intriguing that a dark matter candidate with a mass of a few GeV would have an abundance 
similar to that of ordinary matter \cite{Kaplan:1991ah,Barr:1991qn,Kitano:2004sv,Dodelson:1991iv,Farrar:2005zd,Cosme:2005sb,Kaplan:2009ag}. 
Also, there is potentially a whole zoo of light WIMP candidates~\cite{Feng:2008dz} 
which are yet not constrained but which are within reach of existing or forthcoming experiments.

The elastic scenario has already been addressed in Ref.~\cite{Andreas:2008xy}, 
following an analysis based on a two bins version of the recent DAMA data of Ref.~\cite{Petriello:2008jj}. 
In Ref.~\cite{Andreas:2008xy} it was shown that the elastic scattering of a light scalar particle interacting dominantly through 
the Higgs channel, may be simultaneously compatible with both the DAMA data and with the WMAP cosmic 
abundance~\footnote{This is also specific to a scalar dark matter candidate \cite{Andreas:2008xy}. For instance, 
it is well-known that annihilation of a fermionic singlet through the Higgs is a P-wave suppressed process. 
The coupling required to explain the relic abundance is then way too large to be compatible with direct 
searches \cite{Andreas:2008xy}. More complicated fermionic models are however viable \cite{Kim:2009ke}.}. 
This limit of the IDM is also the simplest instance of a Higgs portal \cite{Patt:2006fw} and so has a larger scope. 
A singlet scalar as a candidate for dark matter has been discussed in various works \cite{McDonald:1993ex,Burgess:2000yq,Barger:2007im}.  
\begin{figure}[t!]
\centering
\includegraphics[width=0.5\columnwidth]{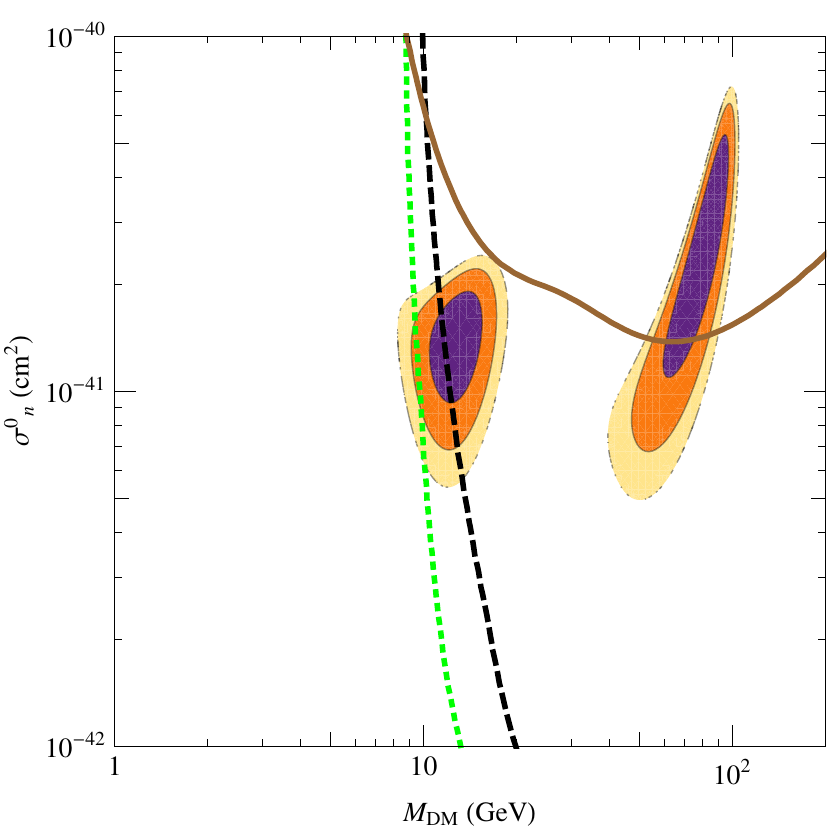}
\caption{\underline{Elastic scenario}\ \ (model independent, $f_p=f_n$): Allowed regions in the plane $\sigma^0_{n} - M_{DM}$ consistent with the DAMA annual modulation 
signal at $90\%$, $99\%$ and $99.9\%$ CL. The solid (brown) curve shows the total unmodulated rate for DAMA. The escape 
velocity is $v_{esc} = 650\  \rm{km/s}$. The dashed (black) curve is the exclusion contour at $99\%$ CL for CDMS-Ge, 
while the dotted (green) curve is the exclusion limit at $99\%$ for XENON10.}
\label{fig:mlbda_ind}
\end{figure}
\begin{figure}[t!]
\begin{minipage}[t]{0.5\textwidth}
\centering
\includegraphics[width=0.985\columnwidth]{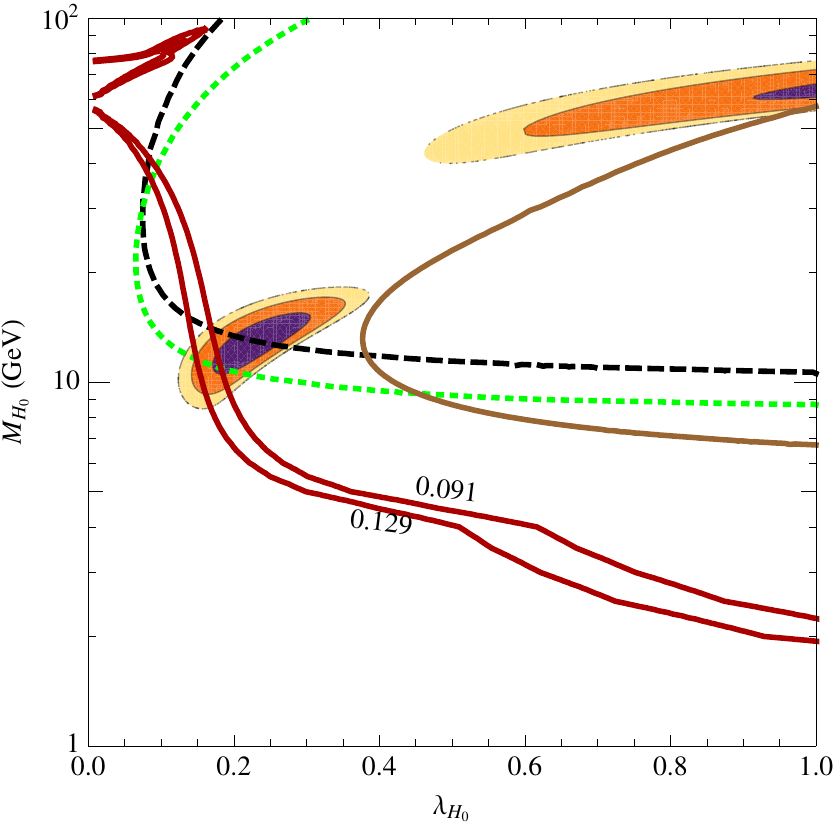}
\end{minipage}
\hspace*{-0.2cm}
\begin{minipage}[t]{0.5\textwidth}
\centering
\includegraphics[width=0.985\columnwidth]{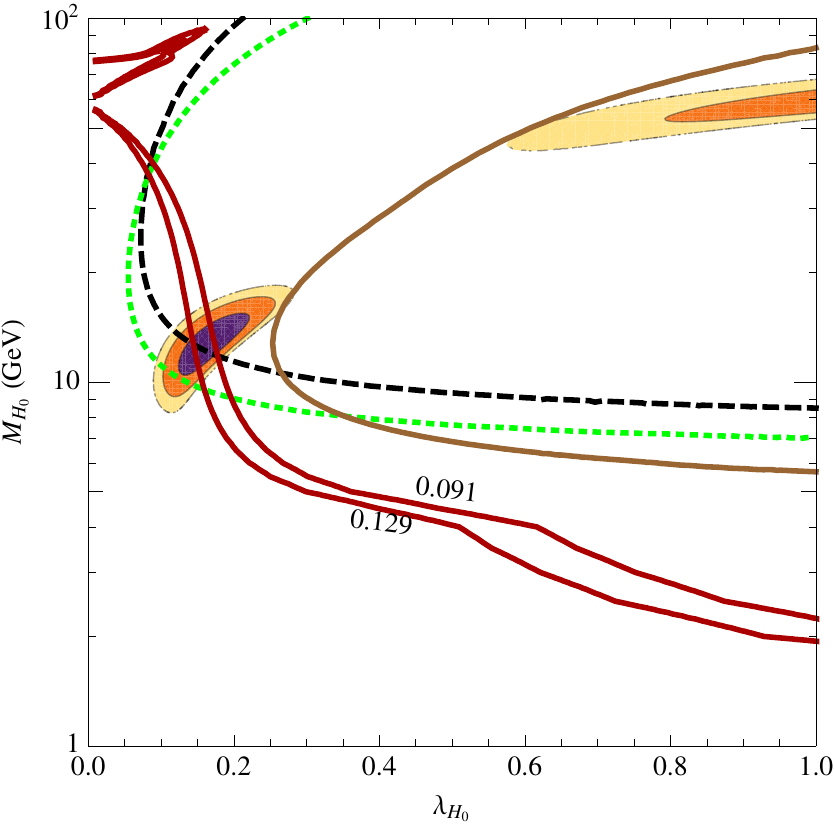}
\end{minipage}
\caption{\underline{Elastic scenario (predictions of IDM)}\ \ Left panel for $v_{esc}=450$ km/s and right panel for $v_{esc}=650$ km/s. Allowed regions in the plane $|\lambda_{H_0}| - M_{H_0}$ consistent with the DAMA annual modulation signal at $90\%$, $99\%$ and $99.9\%$ CL. 
The red (solid) lines delimit the WMAP 5 years bounds at $3\ \sigma$ CL~\cite{Komatsu:2008hk}.
The solid (brown) curve shows the total unmodulated rate for DAMA. The dashed (black) curve is the exclusion contour at $99\%$ CL for CDMS-Ge, while the dotted (green) curve is the exclusion limit at $99\%$ for XENON10.}
\label{fig:mlbda}
\end{figure}

The model independent analysis of Ref.~\cite{Petriello:2008jj}, which was based on DAMA data grouped in two bins, 
$2-6$ keVee and $6-14$ keVee gave a SI cross section in the range
\begin{equation}
\label{eq:range_s}
3 \times 10^{-41}\, \mbox{\rm cm}^2 \lsim \sigma_n^{SI} \lsim 5 \times 10^{-39} \, \mbox{\rm cm}^2
\end{equation}
and a dark matter mass in the range
\begin{equation}
\label{eq:range_m}
3 \, \mbox{\rm GeV} \lsim M_{DM} \lsim 8 \, \mbox{\rm GeV},
\end{equation}
where the upper bound on the mass is set by the exclusion limit from XENON. 
Subsequently, various groups have emphasized the relevance of using the full set of spectral data provided by 
DAMA \cite{Chang:2008xa,Fairbairn:2008gz,Savage:2008er,Savage:2009mk}. These spectral data are shown in Figure~\ref{fig:chi_el},  
together with the theoretical spectrum for the elastic scattering of a $12$ GeV candidate. 
In the IDM, this would correspond to a light  scalar  with coupling $\vert \lambda_{H_0}\vert = 0.24$  
to the Higgs, and $M_h = 120$ GeV. The raise at small recoil energies is typical of elastic scattering of a light WIMP. 

\begin{figure}
\centering
\includegraphics[width=0.6\columnwidth]{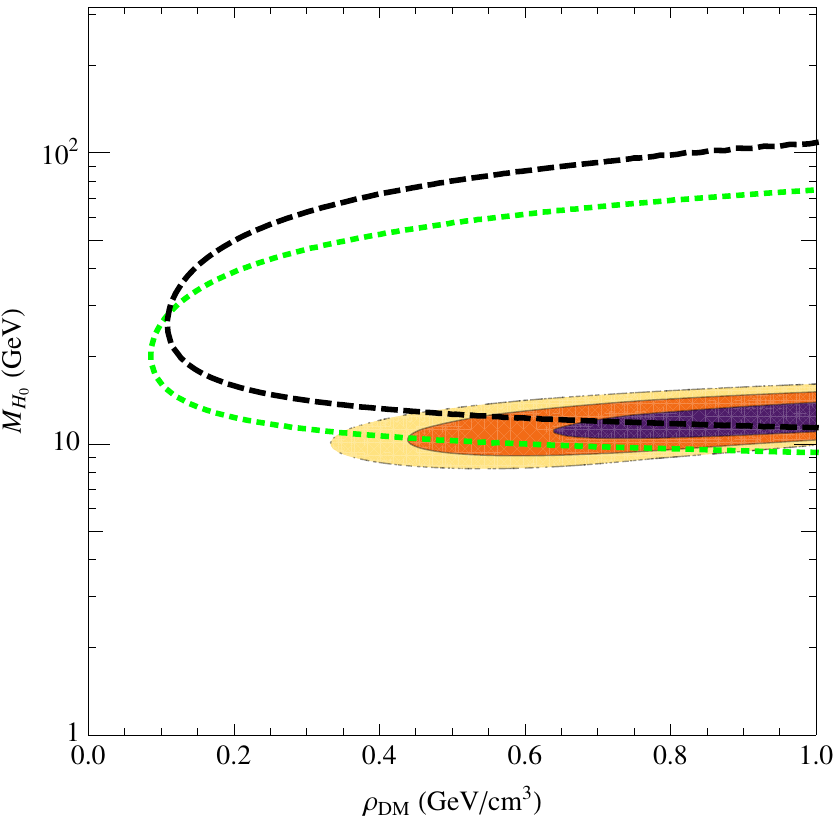}
\caption{\underline{Elastic scenario (predictions of IDM)}\ \ Allowed regions in the plane $\rho_{\rm{DM}} - M_{H_0}$ consistent with the DAMA annual modulation signal at 
$90\%$, $99\%$ and $99.9\%$ CL for a fixed coupling $|\lambda_{H_0}| = 0.12$. The escape velocity is fixed to $v_{esc} = 650\  \rm{km/s}$. The bounds are the same as in Figure~\ref{fig:mlbda}.}
\label{fig:mrho}
\end{figure}

When the full spectral information of DAMA is used, the allowed region of parameters is much reduced compared to 
the ranges of Eqs.~(\ref{eq:range_s}) and (\ref{eq:range_m}). Our main results regarding a light WIMP in the IDM 
are summarized in Figures \ref{fig:mlbda_ind} and \ref{fig:mlbda} . 
Figure  \ref{fig:mlbda_ind} is model independent, the only hypothesis being that the coupling to nuclei may be approximated with $f_n=f_p$. 
It shows the region allowed by DAMA, up to $99.9\%$ CL, together with the exclusion limits from XENON-10 and CDMS-Ge, 
which are the only two experiments relevant in these ranges.  Our results are consistent 
with those of Refs.~\cite{Chang:2008xa,Fairbairn:2008gz,Savage:2008er,Savage:2009mk}. 
A small region of masses and cross sections  is compatible with all experiments, albeit only at the $3 \sigma$ level. 
Figure \ref{fig:mlbda} gives the allowed parameter space for the IDM, or equivalently for a light scalar singlet, the left panel is for an escape velocity of 450 km/s and the right one for $v_{esc}=650$ km/s.
The mass of the Higgs has been fixed to $M_h =120$ GeV, but all the predictions only depend on the ratio $\lambda_{H_0}^2/M_h^4$ 
and may thus be trivially rescaled~\cite{Andreas:2008xy}. Also shown in the figure is the predictions for 
the abundance from thermal freeze-out, compared with the WMAP data, $0.091 \leq \Omega_{DM} h^2 \leq 0.129$ at $3 \sigma$ \cite{Chang:2008xa}. 
We should perhaps emphasize the fact that no further adjustment of parameters is done in order to fix the relic abundance.

The overlap between the  region allowed by DAMA and the WMAP abundance is amusing, but should perhaps not be taken too seriously, 
independently of the fact that this region is actually excluded by the other experiments. 
One may envision at least three ways in which the regions of parameters may be modified. 
First the local abundance is fixed here to the fiducial value $\rho_{DM} = 0.3$ GeV/cm$^3$. 
Decreasing the local abundance will move the DAMA region and the excluded region to larger quartic couplings, 
without affecting the cosmic abundance. See also Figure \ref{fig:mrho}, where we show this variation for a fixed quartic coupling. 
Second, one should take into account the fact that the coupling of the Higgs to a nucleon $N$, or $g_{hNN}$, is not well-known. 
This is parameterized by a form factor $f$ using the trace anomaly, 
$f m_N \equiv \langle N\vert \sum_q m_q \bar q q\vert N \rangle= g_{hNN} v$. The reference value for 
$f=0.3$ \cite{Pavan:2001wz,Bottino:1999ei,Koch:1982pu,Gasser:1990ap}, but it may vary within a quite large 
range $0.15 \leq f \leq 0.6$~\cite{Andreas:2008xy}. 
Changing the parameter $f$ will move around the DAMA and excluded region, 
without affecting the relic abundance. Finally, one may modify the relic abundance by changing the cosmology. 
\section{Heavier WIMPs and inelastic scattering}\label{sec:inelwimp}

\begin{figure}
\begin{minipage}[t]{0.65\textwidth}
\centering
\includegraphics[width=0.95\columnwidth]{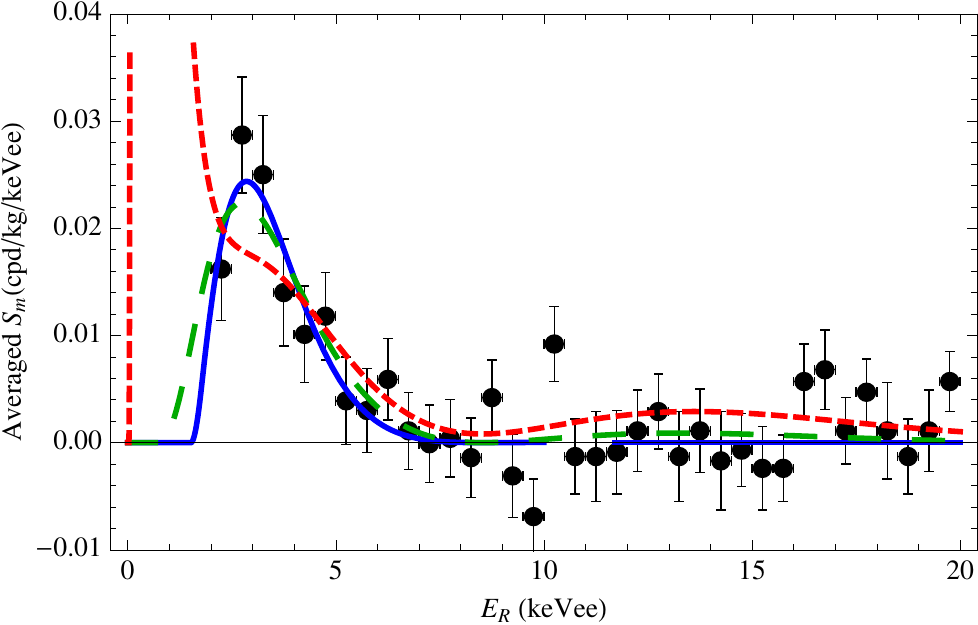}
\end{minipage}
\hspace*{-0.2cm}
\begin{minipage}[t]{0.35\textwidth}
\vspace*{-6cm}
\centering
\begin{tabular}{|cc|}
\hline
\includegraphics[width=0.25\columnwidth]{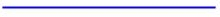}  & $M_{H_0}= 69$ GeV\\ 
& $\delta= 136$ keV \\
&  $\chi^2 = 32.7$  \\ 
& \\
\includegraphics[width=0.25\columnwidth]{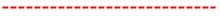}  & $M_{H_0}= 3840$ GeV\\
& $\delta= 124$ keV \\
&  $\chi^2 = 32.7$  \\ 
& \\
\includegraphics[width=0.25\columnwidth]{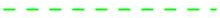}  & $M_{H_0}= 2206$ GeV\\
& $\delta= 26$ keV \\
&  $\chi^2 = 36.8$  \\ 
\hline
\end{tabular}
\end{minipage}
\caption{\underline{Inelastic scenario (predictions of IDM)}\ \ On the left, the averaged modulated amplitude $S_{m}$ for the best fit points (for $v_{esc}= 600$ km/s, see also Figure~\ref{fig:mdelta}). 
On the right: Values for the global and local minima of the chi-square distribution for the best fit points. 
The $\chi^2_{min}$ is evaluated for 34 d.o.f. .}
\label{fig:chi_inel}
\end{figure}

In the inelastic Dark Matter scenario, or iDM for short, an incoming dark matter particle $DM_1$ from the halo is supposed 
to scatter with a nucleus into a slightly heavier state $DM_2$, with a mass splitting $\delta = M_{DM_2} - M_{DM_1} \sim 100$ keV. 
In this scenario, which has been first proposed in Ref.~\cite{TuckerSmith:2001hy} and confronted to the recent data  
in Refs.~\cite{TuckerSmith:2004jv,Chang:2008gd,MarchRussell:2008dy,Cui:2009xq,Finkbeiner:2009ug}, 
a much broader range of dark matter candidates may both fit DAMA and be consistent with the other experiments. 
To the naked eye the fit to the modulated spectrum is better than in the elastic case, see Figure~\ref{fig:chi_inel}, 
although the GOF to data is comparable. 
On one hand, the iDM scenario opens the exciting possibility that there might be in the dark sector more than just one plain particle. 
On the other hand, from the point of view of particle physics, the naturalness of the small mass splitting needs to be addressed \cite{Cui:2009xq}.

\begin{figure}
\begin{minipage}[t]{0.5\textwidth}
\centering
\includegraphics[width=\columnwidth]{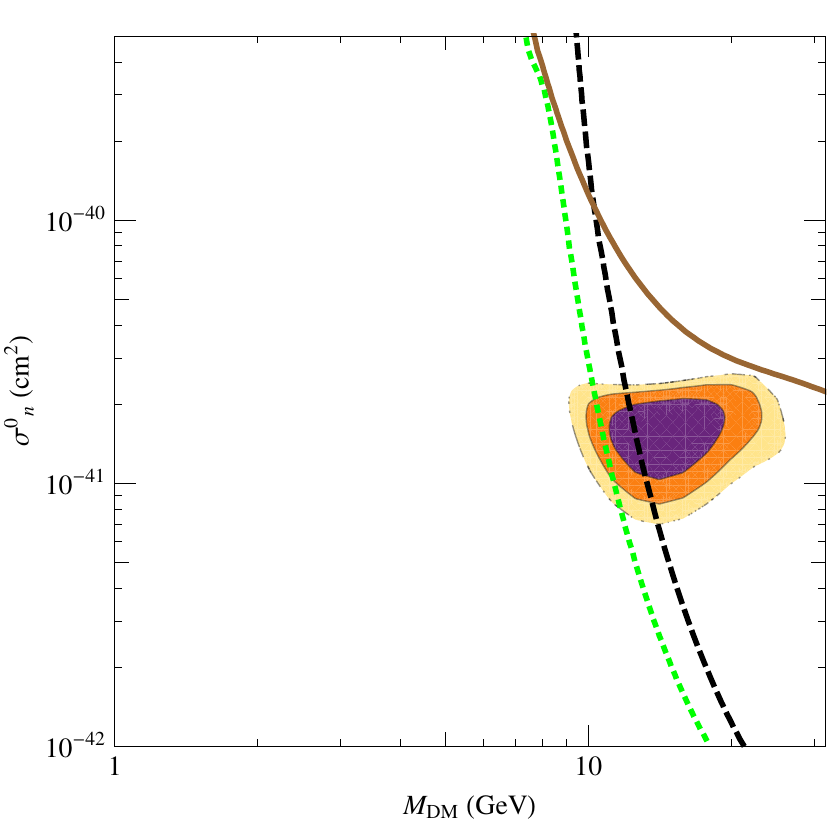}
\end{minipage}
\hspace*{-0.2cm}
\begin{minipage}[t]{0.5\textwidth}
\centering
\includegraphics[width=0.985\columnwidth]{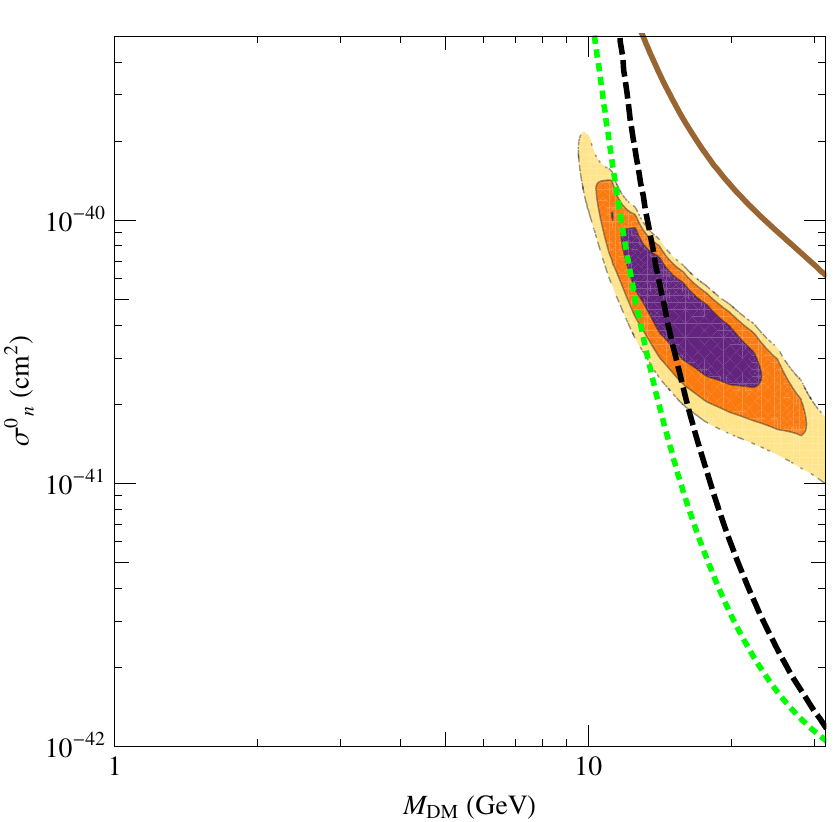}
\end{minipage}
\\
\begin{minipage}[t]{0.5\textwidth}
\centering
\includegraphics[width=\columnwidth]{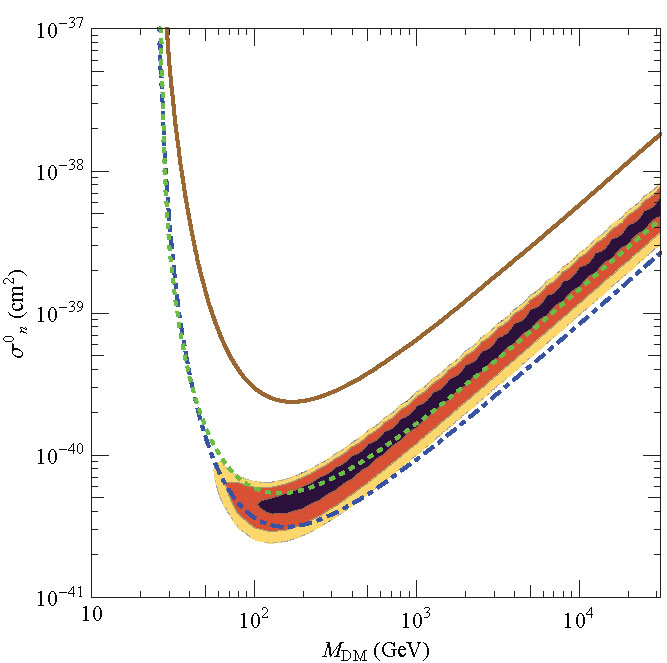}
\end{minipage}
\hspace*{-0.2cm}
\begin{minipage}[t]{0.5\textwidth}
\centering
\includegraphics[width=0.985\columnwidth]{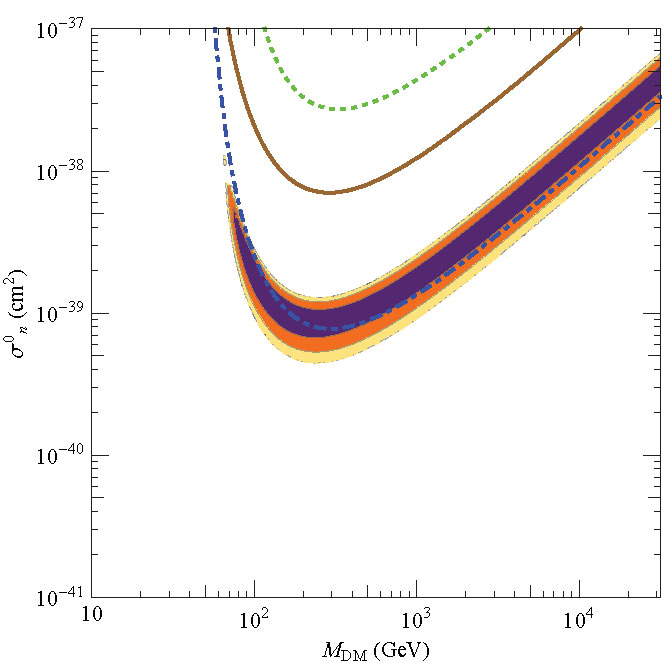}
\end{minipage}
\caption{\underline{Inelastic scenario (model independent, $f_p=f_n$)}\ \ 
Allowed regions in the plane $M_{DM} - \sigma^0_{n}$ consistent with the DAMA annual modulation signal at $90\%$, $99\%$ and $99.9\%$ CL 
for different mass splittings. From left to  right and from top to bottom, respectively, 
$\delta = 5$ keV, $\delta = 20$ keV, $\delta= 80$ keV and $\delta= 150$ keV. The escape velocity is $v_{esc} = 650\  \rm{km/s}$.  
The total unmodulated rate expected at DAMA is given by the solid (brown) line. The dotted (green) curve is the exclusion contour at $99\%$ CL for Xenon10, the dashed (black) curve is the upper 
bound coming from CDMS-Ge, while the dotted-dashed (blue) curve is the exclusion limit at $99\%$ for CRESST-II.}
\label{fig:mdm}
\end{figure}

To illustrate the relevance of the inelastic scenario,  we show in Figure~\ref{fig:mdm} four snapshots of the $M_{DM}-\sigma^0_n$ 
plane for different mass splittings, using Eq.(\ref{cross_section}) with $f_p=f_n$ with increasing splittings. 
These panels show that, as the mass splitting is increased  the two regions of Figure~\ref{fig:mlbda} 
(which correspond to zero mass splitting) merge and move toward higher dark matter masses and larger cross sections, and more solutions open.  

The possibility of dark matter from a weak  $SU(2)$ doublet  has already been considered in the 
literature \cite{TuckerSmith:2004jv,Chang:2008gd,Cui:2009xq}.  
Here we focus on the prediction of the IDM, which is the only model among Scalar Multiplet Models where a mass 
splitting of the order of $100$~keV,
small enough to be relevant for inelastic scattering, can appear at the renormalizable level. 
Notice that when $\delta = M_{A_0}-M_{H_0} \ll M_{H_0}$, the only accelerator constraint on IDM comes from the $Z$ boson decay width.
In particular, this limit is not excluded by LEPII measurements~\cite{Lundstrom:2008ai}.  
In the IDM, an accidental $U(1)_{PQ}$ symmetry arises in the limit of vanishing $\lambda_5$, so that the mass 
splitting $\delta$ is protected against
radiative corrections. In this sense, the option of having a small quartic coupling $\lambda_5$ is technically natural.\footnote{As put forward in Refs. \cite{Kadastik:2009dj,Kadastik:2009cu}, the Inert Doublet Model may be embedded in SO(10), provided the Standard Model fermions are also odd under $Z_2$ or matter-parity, with the Standard Model in a $\bf 10$ of SO(10) and the inert doublet in a $\bf 16$. A very interesting feature of this framework is that the coupling $\lambda_5$ is absent at the level of renormalizable operators \cite{Kadastik:2009cu}. Higher order operators may break the accidental PQ symmetry while preserving matter-parity (which is a gauged discrete symmetry and thus protected). In Ref. \cite{Kadastik:2009cu}, breaking by quantum gravity effect are estimated to give $\lambda_5 \sim (M_{SO(10)}/M_{\rm Pl})^n$ with $n\geq 1$. For $n=2$ and $M_{SO(10)} \sim 10^{16}$ GeV for the scale of grand unification scale, gives $\lambda_5 \sim 10^{-7}$, in the range required to explain DAMA through inelastic scattering, Eq.(\ref{lambdacinq}). We would like to emphasise that higher operators may also arise within SO(10). For instance, with scalar fields in a $\bf 126$, which is required to break $\bf B-L$, one may build a dimension 5 operator suppressed by some scale $\Lambda$ (at tree level, this may be mediated by scalars in a $\bf 54$), which gives $\lambda_5 \sim v_{126}/\Lambda$ after breaking of $\bf B-L$ (we thank Thomas Hambye for discussions on this point).}

\begin{figure}
\centering
\includegraphics[width=0.6\columnwidth]{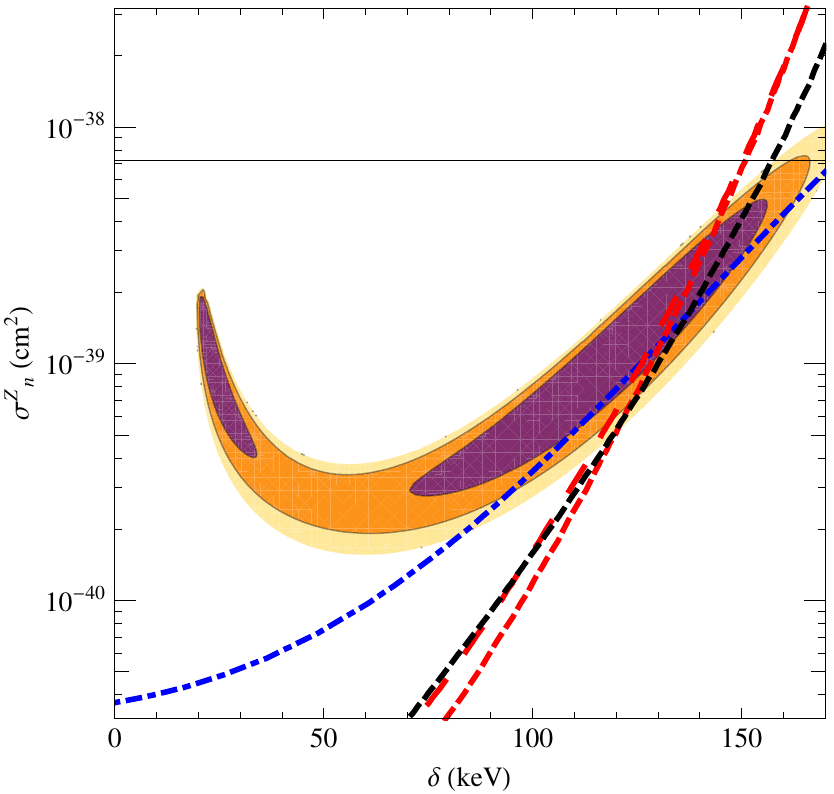}
\caption{\underline{Inelastic scenario (model independent, $f_ p= 4 \sin^2\theta_W-1, f_n= 1$)}\ \ 
Allowed regions in the plane $\delta - \sigma_{n}^{Z}$ consistent with the DAMA annual modulation signal at 
$90\%$, $99\%$ and $99.9\%$ CL for a fixed mass $M_{DM} = 525$ GeV. The escape velocity is $v_{esc} = 600\  \rm{km/s}$.  The dashed (black) curve is the exclusion contour 
at $99\%$ CL for CDMS-Ge, the dotted-dashed (blue) curve is the exclusion limit at $99\%$ for CRESST-II,
the long-dashed (red) line denotes the exclusion limit coming from ZEPLIN-III  
and the short-dashed (red) line is that from ZEPLIN-II. 
The grey solid line corresponds to the $Z$ exchange with a neutron.}
\label{fig:msigz}
\end{figure}

For the sake of comparison with other works, in particular Refs.~\cite{MarchRussell:2008dy,Cui:2009xq} with which we agree, 
we show in Figure~\ref{fig:msigz} the allowed regions in the $\delta-\sigma_n^Z$ plane, where $\sigma_n^Z$ is the 
WIMP cross section with a neutron for $SU(2)$-type couplings  ($f_p =4 \sin^2\theta_W-1$, and $f_n=1$) but 
for an arbitrary scale (the horizontal line at $\sigma_n^Z \approx 7\cdot 10^{-39}$ is the SM $Z$-boson exchange)
~\footnote{Unlike Ref.~\cite{Cline:2009xd}, we find that ZEPLIN-II is not much more constraining than other experiments, like ZEPLIN-III, 
in the region of the inelastic scenario.
This is due to the fact that the DAMA region we have found,
which is consistent with results in Refs.~\cite{MarchRussell:2008dy,Cui:2009xq},
differs from that of Ref.~\cite{Cline:2009xd}.}.

\begin{figure}
\begin{minipage}[t]{0.5\textwidth}
\centering
\includegraphics[width=0.985\columnwidth]{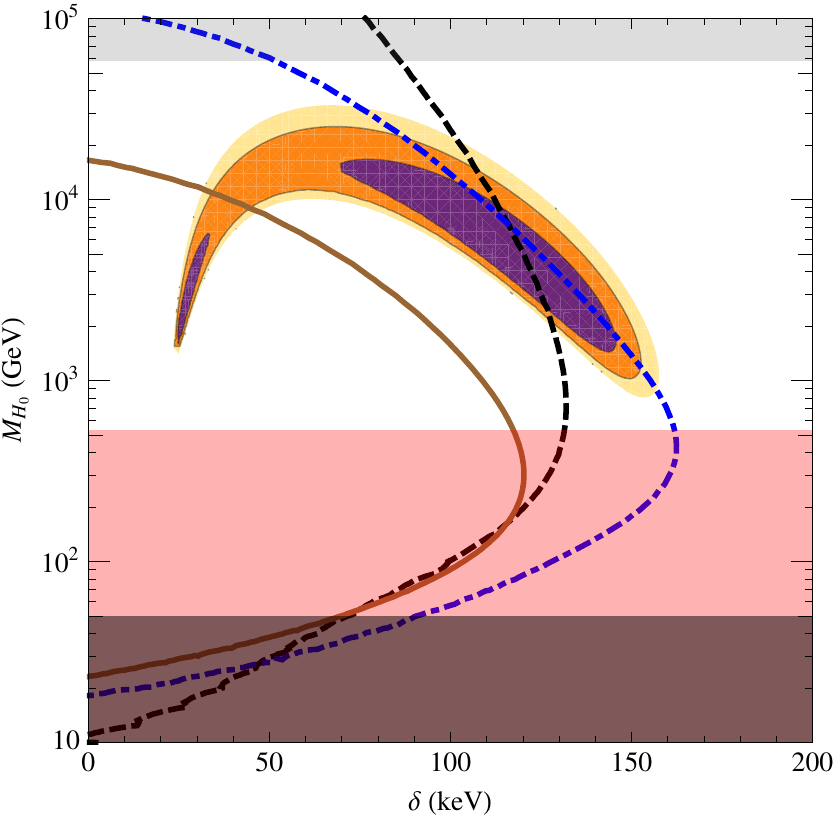}
\end{minipage}
\hspace*{-0.2cm}
\begin{minipage}[t]{0.5\textwidth}
\centering
\includegraphics[width=0.985\columnwidth]{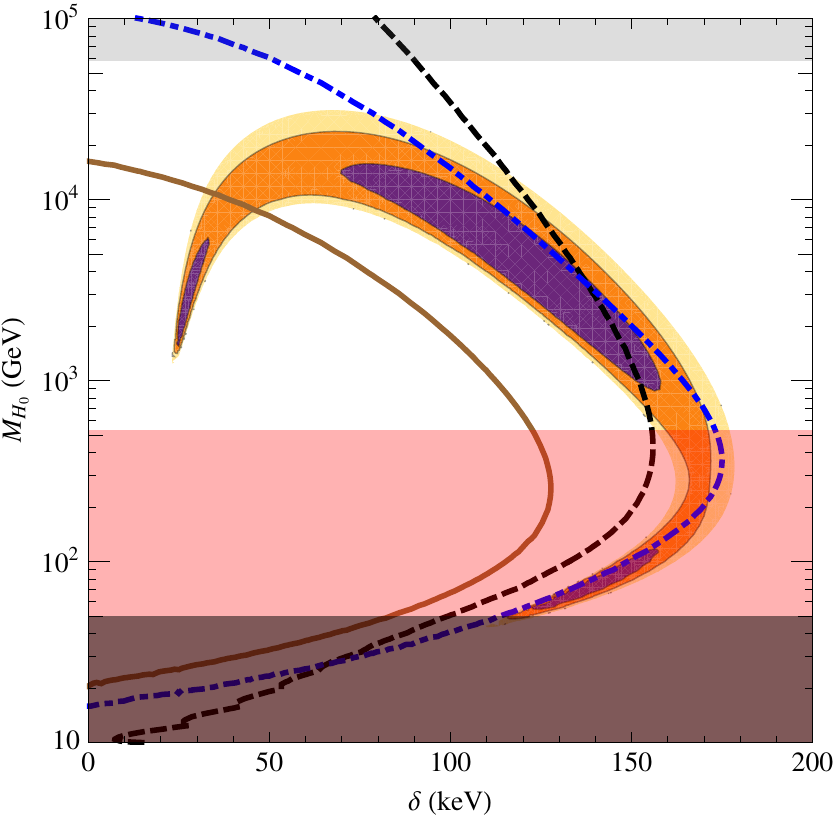}
\end{minipage}
\caption{\underline{Inelastic scenario (predictions of IDM)}\ \ Left panel for $v_{esc}=450$ km/s and right panel for $v_{esc}=600$ km/s.
Allowed regions in the plane $\delta - M_{H_0}$ consistent with the DAMA annual modulation signal at 
$90\%$, $99\%$ and $99.9\%$ CL. The solid (brown) curve shows the total unmodulated rate for DAMA. The dashed (black) curve is the exclusion contour at $99\%$ CL for CDMS-Ge, while the dotted-dashed (blue) 
curve is the exclusion limit at $99\%$ for CRESST-II. The brown (dark grey bottom) region is excluded by LEPI measurement of the width of the $Z$; the pink region (light grey bottom) is below the WMAP abundance; the grey region (top) corresponds to the Griest-Kamionkowski unitarity limit, Refs \cite{Griest:1989wd,Hambye:2009pw}.}
\label{fig:mdelta}
\end{figure}

In Figure~\ref{fig:mdelta}, we show the predictions of the IDM, the left panel is for an escape velocity of 450 km/s and the right one for $v_{esc}=600$ km/s. These figures show that there exists a whole range of candidates, between $M_{H_0} \sim 535$~GeV and $M_{H_0} \sim 20$~TeV, 
which are compatible with DAMA and all the other experiments, and which have a relic abundance consistent with WMAP. The candidates in the pink region (light grey bottom) have an abundance below the WMAP observation. The grey region (top) corresponds to the Griest-Kamionkowski unitarity limit $M_{H_0}< 58$ TeV, see Refs. \cite{Griest:1989wd,Hambye:2009pw}. 
Notice that the cross section in the inelastic dark matter scenario is fixed, as the scattering occurs dominantly through a $Z$ boson. Therefore the relevant parameter space is $M_{H_0}-\delta$. The brown (dark grey bottom) region is excluded by LEPI measurement of the width of the $Z$.

Heavy candidates with a mass above $535$~GeV have been shown to be compatible with WMAP in Ref.~\cite{Hambye:2009pw}, 
irrespective of the precise value of the small splitting $\delta$. The agreement can be achieved by tuning the values of $\mu_2$ and $M_{H^+}$,
which control the amount of coannihilation. For candidates with a mass lower than $535$~GeV, 
direct annihilation channels into $Z$ or $W$ bosons, when opened, 
reduce the relic abundance below the WMAP value.
For candidates lighter than the $W$ threshold, but above the lower bound set by the LEP measurement on the decay width of the Z,
the coannihilation cross section through the $Z$ boson in the $s$-channel is large due to the proximity of the $Z$ pole.
Even at $M_{H_0}=80$~GeV, the total coannihilation cross section is about $25$~pb, much larger than $1$~pb, the typical value needed for WMAP.
Therefore, middle mass range candidates with inelastic interactions that are consistent with DAMA have a small relic abundance.

\begin{figure}
\centering
\includegraphics[width=0.5\columnwidth]{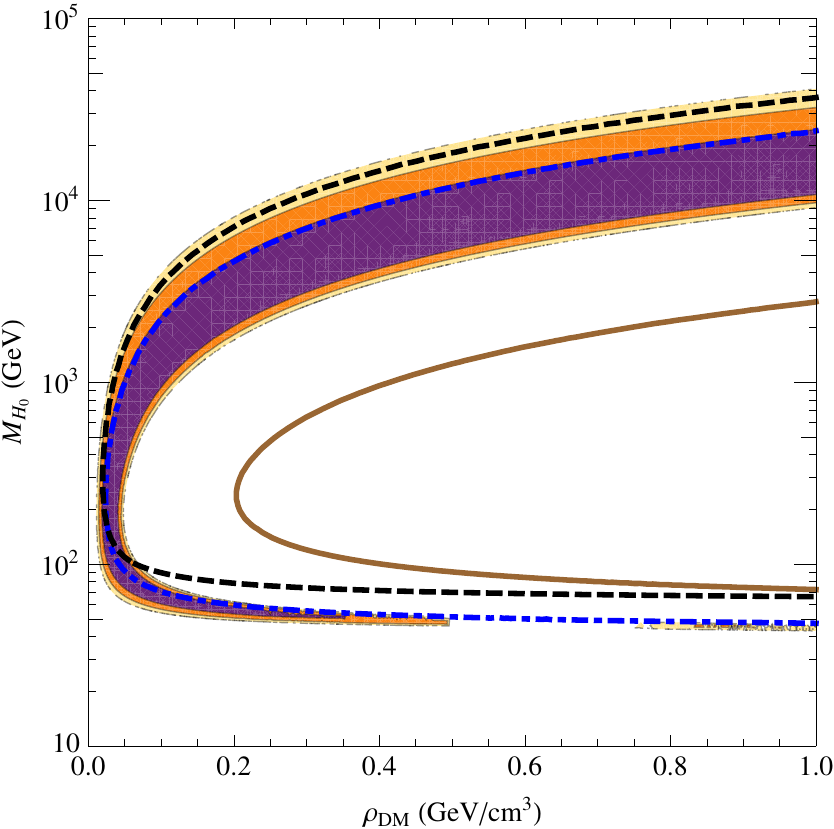}
\caption{\underline{Inelastic scenario (predictions of IDM)}\ \ 
Allowed regions in the plane $\rho_{DM} - M_{H_0}$ consistent with the DAMA 
annual modulation signal at $90\%$, $99\%$ and $99.9\%$ CL for a fixed mass splitting $\delta = 120$ keV. 
The solid (brown) curve shows the total unmodulated rate for DAMA.  
The escape velocity is $v_{esc} = 600\  \rm{km/s}$.  
The dashed (black) curve is the exclusion contour at $99\%$ CL for CDMS-Ge, while the dotted-dashed (blue) 
curve is the exclusion limit at $99\%$ for CRESST-II.}
\label{fig:mrho_inel}
\end{figure}

With respect to this problem, we put forward two options. One possibility is that  the neutral scalar is a 
subdominant component of dark matter and we show how the fit changes as the 
local density of the neutral scalar is changed~\footnote{The local density can also vary due to the clumpy nature of DM in galactic halos. 
The value of $0.3$~GeV$/{\rm cm}^3$ usually quoted only reflects an
average density of the smooth component at the location of the Earth.} in Figure~\ref{fig:mrho_inel}.
Alternatively one may envision the possibility that there is a charge asymmetry in the dark sector. 
Concretely, we may use the approximate PQ symmetry, which becomes exact in the limit $\lambda_5 \rightarrow 0$.  
We would like to argue here that, for small enough $\lambda_5$, an initial PQ charge asymmetry produced in the 
early universe  may survive until after freeze-out. 
Indeed, in the limit $\lambda_5 \rightarrow 0$, the complex, neutral scalar field $H_n \equiv (H_0 + i A_0)/\sqrt{2}$, 
has a PQ charge $+1$ which is conserved by gauge interactions and
the scalar interactions given by the potential Eq.~(\ref{potential}).
For small $\lambda_5$, this charge is only broken by processes proportional to $\lambda_5^2$, in the annihilation channels 
$H_n H_n \rightarrow h h$ or $H_n H_n \rightarrow (h) \rightarrow f \bar{f}$,
and the corresponding scattering processes $H_n X \rightarrow H_n^* X$.  
In full generality, the resolution of the Boltzmann equation that governs the evolution of the comoving charge density
$\Delta=(n_{H_n}-n_{H_n^*})/s(T)$, where $s(T)$ is the entropy density at temperature $T$, are likely to be intricate. 
Details about the electroweak phase transition, like the temperature dependence of the Higgs mass or 
the fate of the Goldstone bosons, can affect the result. Also, the number of processes that violate the PQ charge is temperature dependent.
In particular, the annihilation channels into Higgs, top quark and bottom quark may or may not be present.

To simplify the problem, here we will consider that the contact interactions 
$H_n H_n \xrightarrow{\sigma_A} h h$ (annihilation) and $H_n h \xrightarrow{\sigma_S} H^{\ast}_n h$ (scattering)
are the dominant processes.
This assumption is justified by the fact that channels with an intermediate Higgs like $H_n H_n \rightarrow (h) \rightarrow f \bar{f}$ 
are further suppressed by Yukawa couplings. Also, at high temperature before the freeze-out, $n_{H_n} \simeq n_{H_n^*}$.
With these hypotheses, the evolution of the asymmetry is controlled by a simple Boltzmann equation
\be
\frac{d\Delta}{dx} \approx -2\alpha \left( \gamma_A \, \frac{Y_n}{(Y_n^{eq})^2} + \gamma_S \, 
\frac{Y_h}{Y_n^{eq} Y_h^{eq}} \right) \Delta \quad ,
\ee
with $\alpha = x/H s(T)$, $x = M_{DM}/T$, $H$ is the Hubble parameter, and $Y_i \equiv n_i/s(T)$ is the comoving number density of the species $i$.
The thermal average of a cross-section $\sigma$ for the process $1 2 \leftrightarrow 3 4$ is given by
$\langle \sigma v \rangle = \gamma(1 2 \leftrightarrow 3 4)/n_1^{eq}(T) n_2^{eq}(T)$,
with~\cite{Luty:1992un}
\be
\gamma(1 2 \leftrightarrow 3 4) = \frac{T}{8\pi^4} \int ds \, \sqrt{s} \, K_1(\sqrt{s}/T) \, p_{{\rm 1cm}}^2 \, \sigma(1 2 \leftrightarrow 3 4) \quad .
\ee
For $M_{DM} \rightarrow 0$, the equilibrium number density reduces to
\be
n^{eq} = \frac{g M_{DM}^2 T}{2\pi^2} K_2(M_{DM}/T) \simeq \frac{g T^3}{\pi^2} \quad ,
\ee
with $g$ the number of internal degrees of freedom.

To further simplify, we will neglect the masses of the particles in computing thermal averaged quantities. 
This will provide an upper bound on the relevant interaction rate, as a finite mass can only reduce the number density in the thermal bath.
In this limit, the annihilation and the scattering cross sections are simply given by
\be
\sigma_A = \frac{1}{2} \sigma_S = \frac{\lambda_5^2}{32\pi s} \quad ,
\ee
with $s$ the center of mass energy and the thermal average $\gamma$ is given by
\be
\gamma_A = \frac{\lambda_5^2 T^4}{2^8 \pi^5} \, .
\ee
Using the relation $H = 1.66 \sqrt{g_*} T/M_{Pl}$, the out-of-equilibrium condition for the interaction rate, 
$\Gamma_A = \langle \sigma_A v \rangle n^{eq} < H$, implies
\be
\lambda_5 < 10^{-7} \, g_*^{1/4} \sqrt{\frac{T}{10~\rm{GeV}}} \quad .
\ee
By comparing this relation with Eq.~(\ref{lambdacinq}), we see that PQ breaking processes are indeed out-of-equilibrium in the early universe,
for candidates lighter than about $100$~GeV. For heavier candidates ($M_{DM} \simeq 1 \dots 10~\rm{TeV}$), the charge asymmetry may be washed out,
a thermal relic abundance consistent with WMAP can be obtained through the standard freeze-out mechanism, without invoking an asymmetry.

\section{Conclusions and Prospects}\label{sec:conclusions}

In this article we have confronted the inert doublet model or IDM to DAMA. In this very simple framework, which consists of just one extra, inert, 
Higgs doublet, it is possible to explain the DAMA data in two particular limits of the model. 

In the first limit, which is protected by a custodial symmetry, all, but the lightest, 
scalars are decoupled and the model is essentially equivalent to a singlet scalar extension of the Standard Model in which the dark sector interact 
with the rest of the world through the Higgs channel, {\em aka} the Higgs portal. 
A fit to the DAMA data requires a rather light scalar dark matter candidate, in the multi-GeV range, which undergoes elastic, spin independent 
scattering with nuclei (essentially Iodine) in the detector.  
This scenario, which works if channelling is effective,  is severely constrained by exclusion limits imposed by XENON-10 and CDMS. 
There is also some tension with the fact that the total, unmodulated rate  needs to be quite large, possibly saturating the signal in the detector. 
The analysis presented here is essentially an update of Ref.~\cite{Andreas:2008xy}, were it was shown that, without adjusting any parameter, 
such a model may simultaneously explain the DAMA data and be consistent with WMAP. 
Our more refined analysis, which is based on using the full spectrum data provided by DAMA, concurs, as is shown in Figure~\ref{fig:mlbda}. 
Although this scenario, which encompasses a potentially broad range of models,  is challenged by the other experiments, we would like to re-emphasize 
here that it predicts numerous other signatures. 
This stems both from the larger abundance of a light WIMP, and from  the rather large couplings or cross section required to fit the DAMA data. 
Depending on the abundance at the Galactic center, one may have a rather large flux of gamma rays from dark matter annihilations, 
which is already constrained by EGRET, and should be within the reach of the forthcoming Fermi/GLAST data \cite{Feng:2008dz,Andreas:2008xy}. 
Capture by the Sun and the subsequent annihilation into neutrinos may be constrained by Super-Kamiokande 
data \cite{Savage:2008er,Savage:2009mk,Andreas:2009hj,Feng:2008qn}. 
Annihilations in the Galaxy may produce antimatter, in particular in the form of anti-protons and 
anti-deuterons \cite{Bottino:2008mf,Nezri:2009jd}. Finally, a light WIMP in the form of a scalar coupled to the Higgs would imply the 
Higgs mostly decays into a pair of dark matter particles, with striking consequences for its search at the 
LHC~\cite{Burgess:2000yq,Barger:2007im,Andreas:2008xy}. 
It is also intriguing (albeit puzzling) that in this scenario the abundance of dark matter is similar to that of baryons.

In the second limit of the IDM considered in the present paper, the two neutral scalars of the inert doublet are almost degenerate, 
with a splitting ${\cal O}(100 \ $keV), necessary to explain the DAMA data through the inelastic scattering of the lightest of the partners, 
the so-called inelastic Dark Matter or iDM scenario. 
This limit of the IDM is also protected by a symmetry, a $U(1)_{PQ}$ in this case. 
This implies that the splitting, although fine tuned and requiring a small parameter in the form of a quartic coupling, is technically natural. 
Here we show that this very simple model may explain the DAMA data and be consistent with WMAP data for a whole range of candidates, 
with a mass between $\sim 535$ GeV and $\sim 50$ TeV. This is shown in Figure~\ref{fig:mdelta}, in the mass/mass splitting plane. 
This scenario has a GOF which is comparable to that of the elastic scenario, but it is less constrained  by the other experiments 
even if, as far as we know, it does not have a wealth of other potential signatures, 
expected possibly neutrinos from the Sun \cite{Nussinov:2009ft,Menon:2009qj}, or using specific direct detection 
setups \cite{Finkbeiner:2009ug}. 
An issue with the iDM scenario is that it is not easy to explain the smallness of the mass splitting required to fit the 
data \cite{TuckerSmith:2001hy,Cui:2009xq}. 
In the IDM, there is no fundamental explanation for this smallness, but higher scalar multiplets may do better \cite{Hambye:2009pw}. 
For instance, an inert scalar $SU(2)$ triplet with hypercharge $Y=2$ and  with coupling only to the SM Higgs, 
has two neutral scalars which are degenerate at the level of renormalizable operators. 
A splitting may arise through a dimension 6 operator, and thus is suppressed by a factor $\sim v^2/\Lambda^2$, with $\Lambda \gg v$ a new scale. 
Another interesting possibility opened by the model discussed here is that the small mass splitting may be related to a 
tiny violation of a global $U(1)_{PQ}$ symmetry. 
This implies that, at least for not too heavy candidates, $M_{DM} \lsim 100 $ GeV, an initial $U(1)_{PQ}$ 
asymmetry in the Early Universe may survive and 
not be washed-out before annihilations of dark matter particles actually freeze-out. 
An asymmetry in the dark scalar sector could for instance be generated through leptogenesis, if right-handed neutrinos, 
odd under $Z_2$ are also introduced in the model. 
The addition of right-handed neutrinos is actually a very natural extension of the IDM \cite{Ma:2006km}, 
which might also explain the origin of the SM neutrino masses as 
being due to radiative corrections. 
There is some tension between the fact that leptogenesis requires rather heavy right-handed neutrinos, $M_N \sim 10^{11}$ GeV, 
while in the model we have in mind, the SM neutrino masses are  expected to be 
$$
m_\nu \propto {\delta \cdot M_{DM}\over M_N}\, ,
$$
with $\delta \sim 100$ keV and $M_{DM} \sim 100$ GeV, 
which {\em a priori} gives a too small prediction for $m_\nu$. This holds if we assume that the heavy right-handed neutrinos have similar masses and similar Yukawa couplings to 
SM neutrinos. As emphasized in~\cite{Hambye:2009pw} this does not have to be the case, and a hierarchical structure of Yukawa 
couplings together with a light right-handed neutrino, $M_N \sim 1$ TeV, might lead both to successful leptogenesis 
and SM neutrino masses in agreement with observations. We leave this interesting possibility for future investigations..

\acknowledgments

We would like to thank Sarah Andreas and Thomas Hambye for help on this article. Our work is supported by the FNRS-FRS, 
the IISN and the Belgian Science Policy (IAP VI-11).


\bibliographystyle{JHEP}
\bibliography{biblio}

\end{document}